\documentclass[aps,prl,reprint]{revtex4-1}
\usepackage{amssymb,amsmath}
\usepackage{enumitem}
\usepackage{xcolor}
\usepackage{tikz}
\usepackage{tikz-cd}
\usepackage{subfigure}
\renewcommand\sec[1]{\smallskip\noindent\textit{#1}.\quad}
\renewcommand\eqref[1]{(\ref{#1})}
\newcommand{\cF}{{\mathcal F}}
\newcommand{\cM}{{\mathcal M}}
\newcommand\eps{\epsilon^{\alpha \beta}}
\newcommand\barH{\overline{H}}
\newcommand\tr{\mathrm{tr}\,}
\newcommand\comment[1]{}
\newcommand{\CC}{{\mathbb C}}
\newcommand\capt[1]{\caption{\textsf{#1}}}

\begin{document}
\title{Vacuum Geometry of the Standard Model}
\author{Yang-Hui He}
\email{yang-hui.he@merton.ox.ac.uk}
\affiliation{London Institute for Mathematical Sciences, Royal Institution, London, W1S 4BS, UK}
\affiliation{Merton College, University of Oxford, OX1 4JD, UK}
\author{Vishnu Jejjala}
\email{vishnu.jejjala@gmail.com}
\affiliation{Mandelstam Institute for Theoretical Physics, School of Physics, and NITheCS, University of the Witwatersrand, Johannesburg, WITS 2050, South Africa}
\author{Brent D.\ Nelson}
\email{b.nelson@northeastern.edu}
\affiliation{Department of Physics, Northeastern University, Boston, MA 02115, USA}
\author{Hal Schenck}
\email{hks0015@auburn.edu}
\affiliation{Department of Mathematics, Auburn University, Auburn, AL 36849, USA}
\author{Michael Stillman}
\email{mike@math.cornell.edu}
\affiliation{Department of Mathematics, Cornell University, Ithaca, NY 14853, USA}

\begin{abstract}
Vacuum structure of a quantum field theory is a crucial property.
In theories with extended symmetries, such as supersymmetric gauge theories, the vacuum is typically a continuous manifold, called the vacuum moduli space, parametrized by the expectation values of scalar fields.
Starting from the R-parity preserving superpotential at renormalizable order, we use Gr\"obner bases to determine the explicit structure, as an algebraic variety, of the vacuum geometry of the minimal supersymmetric extension of the Standard Model.
Gr\"obner bases have doubly exponential computational complexity (for this case, $7^{2^{1023}}$ operations); we exploit symmetry and multigrading to render the computation tractable.
This geometry has three irreducible components of complex dimensions $1$, $15$, and $29$, each being a so-called rational variety.
The defining equations of the components express the solutions to F-terms and D-terms in terms of the gauge invariant operators and are interpreted in terms of classical geometric constructions.
\end{abstract}

\maketitle

\sec{Introduction}
Insights about an object are often obtained by viewing it in a broader context, that is, by studying its extensions and generalizations.
For example, properties of the real numbers are illuminated by treating them as a subset of the complex numbers.
In physics, over the past half century, profound realizations about quantum field theory (QFT) --- the indisputable theoretical underpinning of the microscopic world --- have come from extensions in dimension (such as string theory and holography) and extensions in symmetry (such as supersymmetry).
One of the most fundamental properties of any QFT is its vacuum, from which all particle excitations arise via application of creation operators.
For the most important QFT in nature, the Standard Model, the vacuum is deduced from the potential of the only scalar field present, the Higgs boson, $\phi$, for which $V(\phi) = -a\phi^2 + b\phi^4$.
As $\phi$ is a complex scalar, by inspection, there is a moduli space of vacua given by $|\phi| = \sqrt{a/2b}$.
The vacuum geometry is therefore topologically a circle $S^1$.
When the electroweak symmetry is spontaneously broken via the Higgs mechanism~\cite{Higgs:1964pj,Englert:1964et,Guralnik:1964eu}, we have chosen a direction in $S^1$, and the vacuum geometry trivializes to a point.

One thing that extended QFTs --- such as supersymmetric (SUSY) gauge theories and string theories --- have taught us is that non-trivial vacuum geometry provides a wealth of information.
As prominent examples, Calabi--Yau structures emerge from string compactification~\cite{candelas1985vacuum} and the Witten index~\cite{witten1982constraints} for SUSY sigma models captures the Euler number of the manifold.
That the Standard Model has something as simple as a circle for its vacuum geometry invites us to look for a natural extension with a more intricate geometrical structure.

The minimal supersymmetric Standard Model (MSSM) is an $\mathcal{N}=1$ QFT in four dimensions with gauge group $G=SU(3)\times SU(2)\times U(1)$ that echoes the spectrum and interactions of particle physics observed in TeV scale laboratories.
To date, experiments have not discovered any physics beyond the Standard Model, so the justification for studying the MSSM, which includes a superpartner for every particle observed in nature, is purely theoretical.
SUSY, for instance, provides an explanation for why radiative corrections to the Higgs mass are small.
The MSSM also supplies a testbed for model building from a fundamental theory of quantum gravity.
Although no construction exactly reproduces the masses of elementary particles and their couplings, there have been numerous top down realizations of MSSM-like chiral theories from superstring theory, \textit{e.g.}, via heterotic compactification~\cite{Braun:2005ux,Bouchard:2005ag}.
For our present purpose, it is therefore natural to ask:
\[
\mbox{What is the vacuum geometry of the MSSM?}
\]

A parallel avenue for elucidating the origin of low-energy physics is to work from the bottom up, beginning with an effective field theory and using the machinery of strings and branes to engineer this structure from ten dimensions.
To proceed, we take inspiration from~\cite{Douglas:1996sw}, in which a SUSY QFT is the worldvolume gauge theory on a D$3$-brane at a special point, and its vacuum moduli space (VMS) gives a geometry that probes the transverse directions to the brane, for instance, telling us how an orbifold is resolved.
The simplest examples are the conifold and theories whose matter content is encoded by quivers that are the affine Dynkin diagrams associated to ADE singularities.
In a series of papers~\cite{Gray:2005sr,Gray:2006jb,Gray:2008yu,He:2014loa,He:2014oha,He:2015rzg}, we have examined aspects of the Higgs branch vacuum geometry of the MSSM with an R-parity preserving renormalizable superpotential.
We consider a three generation model.
There are $49$ component superfields in multiplets labeled $\{Q,u,d,L,e, H, \overline{H}\}$ (or $52$ once we include the right-handed neutrinos $\nu$).

The vacuum geometry is described in terms of the minimal list of gauge invariant operators (GIOs).
There are $973$ ($976$ with right-handed neutrinos included) such operators, which occur in $28$ (or $29$ with neutrinos) types~\cite{Gherghetta:1995dv}.
We list a sampling of these below and refer to~\cite{Gherghetta:1995dv,Gray:2006jb,He:2014oha} for the complete table.
\begin{table}[h]
{\small
\begin{tabular}{|c||c|c|}\hline
\mbox{Type} & \mbox{Explicit Sum} & \mbox{Number} \\
\hline \hline
$QdL$ & $Q^i_{a, \alpha} d^j_a L^k_\beta \eps$ & 27 \\ \hline
$udd$ & $u^i_a d^j_b d^k_c \epsilon^{abc}$  & 9  \\ \hline
$uude$ & $u^i_a u^j_b d^k_c e^l \epsilon^{abc}$  & 27 \\ \hline
$LLe$ & $L^i_\alpha L^j_\beta e^k \eps$  & 9  \\ \hline
$dddLL$ & $d^i_a d^j_b d^k_c L^m_\alpha L^n_\beta \epsilon^{abc} \epsilon_{ijk} \eps$ & 3 \\ \hline
$uuuee$ & $u^i_a u^j_b u^k_c e^m e^n \epsilon^{abc} \epsilon_{ijk}$ & 6 \\ \hline
$uudQdQd$ & $u^i_a u^j_b d^k_c Q^m_{f, \alpha}d^n_f Q^p_{g, \beta} d^q_g \epsilon^{abc} \eps$ & 324 \\ \hline
\vdots& \vdots & \vdots \\ \hline

\end{tabular}}
\capt{A sample of the $28$ GIO types.}
\end{table}

To determine the vacuum, we must solve the F-term and D-term constraints of the theory.
There is an F-term for each field that appears in the superpotential and a D-term for each generator of the gauge group that depends upon the scalar components of chiral fields that are charged with respect to that generator.
The vacuum is obtained from a ring map that implements a symplectic quotient on the space of solutions to the F-terms~\cite{Gray:2009fy,Hauenstein:2012xs}.

Because it is a formidable computational task, to date this question has only been studied for the electroweak sector~\cite{Gray:2005sr,Gray:2006jb,He:2014loa,He:2014oha,He:2015rzg} where the vaccum expectation values (vevs) of scalar quarks are set to zero \textit{ab initio}, and in the SQCD sector~\cite{Gray:2008yu}.
In the former case, we obtain the Veronese surface, a $\mathbb{P}^2$ embedded in $\mathbb{P}^5$ by the complete linear system of conics, as the vacuum geometry, whereas in the latter case, we find that the vacuum geometry is a Calabi--Yau manifold for arbitrary $N_c$ and $N_f$.
In this work, we determine the vacuum geometry of the full MSSM instead of only its subsectors.
The vacuum moduli space consists of three components: a line together with rational varieties of dimensions $15$ and $29$.

\sec{The MSSM superpotential}
On physical grounds, we work with the MSSM fields with renormalizable, R-parity preserving superpotential:
\begin{eqnarray}
W_{\text{minimal}} &=& \; C^0 H_\alpha \barH_\beta \eps + C^1_{ij} Q^i_{a,\alpha} u^{j,a} H_\beta \eps \label{eq:minimal} \\
&& + C^2_{ij} Q^i_{a,\alpha} d^{j,a} \barH_\beta \eps + C^3_{ij} e^i L^j_{\alpha} \barH_\beta \eps ~, \nonumber
\end{eqnarray}
where $i,j=1,2,3$ is a flavor index, $\alpha,
\beta=1,2$ is a $SU(2)$ index, and $a=1,2,3$ is a $SU(3)$ index.
The fields have their usual labels and charge assignments under $G$.
The couplings of matter to the Higgs fields $H$ and $\barH$ generate the masses of the elementary particles.

We assume that the coupling matrices $C$ are \textit{generic}, in the sense that they are invertible.
It is shown in~\cite{HJNSS:2025} that this allows a deformation so that the $C$ are identity matrices.
The singlet neutrinos have both Majorana and Dirac mass terms:
\begin{equation}
W_{\text{neutrinos}} = C^4_{ij} \nu^i\nu^j + C^5_{ij} \nu^i L^j_{\alpha} H_\beta \eps ~. \label{eq:neut}
\end{equation}
The MSSM superpotential $W_{\text{MSSM}}$ is the sum of~\eqref{eq:minimal} and~\eqref{eq:neut}.
Once we include the kinetic terms from the K\"ahler potential, upon SUSY breaking, we recover the terms in the Standard Model Lagrangian.
We now discuss how to solve for the vacuum corresponding to $W_{\text{minimal}}$ and close with the analogous result for $W_{\text{MSSM}}$.

\sec{Methodology}
To review the technique for solving the F-term and D-term equations for the vacuum, let us recapitulate the simplest example.
The worldvolume QFT on a stack of $N$ D$3$-branes is four-dimensional
$\mathcal{N}=4$ super-Yang--Mills theory with gauge group $U(N)$;
it has the superpotential
\begin{equation}
W_{\mathcal{N}=4} = \tr [\Phi_1,\Phi_2]\Phi_3 ~,
\end{equation}
where the $\Phi_i$ are adjoint superfields and the GIOs are traces of products of the $\Phi$s modulo trace relations.
The F-terms tell us that $[\phi_i,\phi_j]=0$, meaning that the scalars $\phi_i \subset \Phi_i$ are simultaneously diagonalizable.
The D-terms do not add additional constraints, so the vacuum geometry is $\mathbb{C}^{3N}/S_N$, corresponding to the diagonal entries of three $N\!\times\!N$ matrices, where we have modded out by the permutation symmetry since the branes are indistinguishable.
This is consistent with the fact that the transverse space to D$3$-branes in a flat ten-dimensional spacetime is $\mathbb{C}^3$.

\sec{An algorithm for the vacuum}
A theorem rediscovered several times~\cite{Buccella:1982nx,Procesi:1985hr,Gatto:1986bt,Luty:1995sd} establishes that the (classical) \textit{vacuum moduli space} is the quotient
\begin{equation}
  {\cal M} = {\cal F}/G^c ~,
    \label{eq:vms}
\end{equation}
where ${\cal F}$, dubbed the \textit{master space}~\cite{Forcella:2008bb,Forcella:2008eh}, is the space of solutions to the partials of the superpotential with respect to the scalar components $\phi$ of the fields $\Phi$, \textit{viz.}, ${\cal F} := \frac{\partial W}{\partial \phi_i}$, and $G^c$ is the complexified gauge group.

From the standpoint of algebraic geometry, this means that our task is to find the defining equation(s) for an object given parametrically.
This is  nontrivial, and is an active research topic in geometric modeling, see Chapter~5 of~\cite{cox:2020}.
A toy example illustrates the process (but masks the difficulty): map $\CC^2 \rightarrow \CC^3$ via $(x,y)\mapsto (x^2, xy, y^2)=:(a_1,a_2,a_3)$.
(We can imagine a QFT where $x$ and $y$ are commuting fields that carry a $\mathbb{Z}_2$ charge $+1$ so that the GIOs are quadratic in the fields.)
The defining equation vanishing on the image is the conic $a_2^2-a_1a_3$.
Our task is to perform the analogous computation when the source space is $\cF \subseteq \CC^{49}$ and there are $973$ of the $a_i$s, which are of degree up to seven, some with scores of monomial terms. 

Formally, the algorithm to determine the vacuum moduli space runs as follows~\cite{Gray:2009fy,Hauenstein:2012xs}:
\begin{enumerate}[itemsep=\parskip,leftmargin=*]\setcounter{enumi}{-1}
\item The input is the action $S$, from which we read off the chiral superfields $\Phi_i$, the gauge group $G$, and the polynomial superpotential $W$.
\item From representations $R_i$ under which the $\Phi_i$ transform, construct a minimal list of gauge invariants, $\{O_j\}$. Here, $i=1,\ldots,m$ and $j=1,\ldots,n$.
\item In the polynomial ring $R = \mathbb{C}[\phi_1,\ldots,\phi_m,y_1,\ldots,y_n]$, (where the $y_j$ are auxiliary variables), construct the ideal $J = \big\langle \partial_i W, y_j - O_j \big\rangle$.
\item Use Gr\"obner bases to eliminate the variables $\phi_i$ from $J \subset R$. This yields the ideal $I_\mathcal{M} \subset \mathbb{C}[y_1,\ldots,y_n]$; vanishing of $I_{\mathcal M}$ defines the vacuum moduli space $\mathcal{M}$.
\end{enumerate}

In highlighting the question at hand, the authors of~\cite{Gherghetta:1995dv} remark that Gr\"obner basis methods are of no practical use for the problem at hand: the Gr\"obner basis algorithm is doubly exponential in the number of variables~\cite{Eisenbud:1995}.
As our source and target spaces have dimensions $49$ and $973$, we are working with $\sim 10^3$ variables. It is natural that~\cite{Gherghetta:1995dv} expressed pessimism.

\sec{Circumventing the computational roadblock}
It turns out that the symmetries present in the system of GIOs result in enough extra leverage that, coupled with the algebraic structure of a multigrading and a careful choice of ordering of the variables (ordering impacts the Gr\"obner basis computation, as we describe in~\cite{HJNSS:2025}), we are able to determine the structure of both the master space $\cF$ and the vacuum moduli space $\cM$. 

The first step is to analyze the structure of the master space $\mathcal{F}$.
In algebraic geometry, an irreducible component corresponds to a prime ideal;
a fundamental result in algebraic geometry is that any variety (zero locus of a system of polynomial equations) can be written as a finite union of irreducible varieties.
This translates into an algebraic operation known as \textit{primary decomposition}: any ideal can be written as a finite intersection of ideals which are almost (up to radical) prime.
As an example, $xy=0$ defines a variety in $\CC^2$, whose irreducible components are the lines $x=0$ and $y=0$. 

In~\cite{HJNSS:2025}, we prove that the master space consists of three irreducible components
\begin{equation}
    \mathcal{F} =
    X_1 \cup X_2 \cup X_3 ~.
\end{equation}
A schematic depicting their dimensions and intersections (which are non-trivial) appears in Figure~\ref{fig:two} below.
Full details on determining these three components and their defining equations appear in \S3 of~\cite{HJNSS:2025}. 

The utility of this decomposition is that rather than analyzing the image of $\cF$ via the map $\phi$, we may instead analyze the images of $X_1,X_2,X_3$ individually.
As the dimensions of these three components are, respectively, $23$, $27$, and $41$ and the gauge group has dimension $12$, the expectation is that the symplectic reductions (the components $M_1, M_2, M_3$ of the image $\phi(\cF) = \cM$) will have dimensions $11$, $15$, and $29$. 

Unexpectedly, $\phi(X_1)=M_1$ has dimension one and is simply a line through the origin in $\CC^{973}$.
The utility of the primary decomposition of $\cF$ comes to the fore in this case: of the $28$ GIO types, only the $9$ GIOs $udd$ are non-vanishing on $X_1$. 

The analysis of $\phi(X_2)=M_2$ is more complicated, but still primary decomposition comes to our aid.
On the component $X_2$, only $5$ of the $28$ GIO types are non-vanishing, comprising a total of $69$ nonzero GIOs.
The key observation is that while these $69$ GIOs do not factor in general, when restricted to $X_2$ they do factor.
We determine that the vanishing ideal of $M_2$ is defined by $6$ linear forms, $816$ quadratics, and $90$ quartics. 

Finally, we come to the analysis of $\phi(X_3)=M_3$.
The ideal of polynomials whose vanishing defines $X_3 \subseteq \CC^{49}$ consists of $4$ linear forms and $4$ quadrics.
However, the hope of a simplification as in the previous two cases is futile: there are $783$ nonvanishing GIOs on $X_3$, in $13$ types.
Analyzing the component $M_3$ requires a new strategy.

The key is to first find equations which define the projection of $M_3$ to a smaller space which preserves the essential structure of $M_3$.
Lengthy computation shows that using only one or two types of GIOs does not yield an image of the expected dimension $29$.
However, we find $81$ different GIO triples which result in the expected dimension.
The GIOs of types $\{QdL, udd, uude\}$ yield defining equations with a clear geometric meaning, and we use those GIOs to obtain a map $X_3\rightarrow Y \subseteq \CC^{63}$.
Note that the $udd$ and $uude$ GIOs correspond to the SUSY neutron and hydrogen atom. 
\begin{figure}[ht]
\begin{center}
  \begin{tikzcd}
    X_3 \arrow[rrr, "\phi"] \arrow[drrr]
    \arrow[dddrrr]
    \arrow[ddddrrr]
    & & & Y_{13} = M_3 \arrow[d, "\pi_{13}"] \arrow[r, hook] & \CC^{783} \arrow[r, hook] & \CC^{973}\\
      & & \vdots & Y_{12} \arrow[d, "\pi_{12}"] \\
      & & & \vdots \arrow[d, "\pi_5"] \\
      & & & Y_{4} \arrow[d, "\pi_4"] \\
      & & & Y_{3} = Y \arrow[r, hook] & \CC^{63}
    \end{tikzcd}
\end{center}
\capt{The map from $M_3$ to $Y$ and liftings.}
\label{ProjectionTower}
\end{figure}

Let $Y_3=Y$ (the subscript denotes the number of GIO types involved).
Our strategy is to add one GIO type at a time to our ``base set'' of the three GIO types $\{uude, udd, QdL\}$, and to show that at each step every one of the ``new'' GIOs can be written as a rational function of the GIOs added previously.
This will establish that each of the maps $\pi_i$ in the vertical chain of maps in Figure~\ref{ProjectionTower} is birational, and consequently so is the chain of compositions. 

Let $Y_4$ be the image of the map obtained using the GIO types $\{uude, udd, QdL, uuuee\}$; there are $6$ GIOs in the $uuuee$ type.
There are $27$ $uude$ GIOs, for simplicity denote them as $a_1,\ldots,a_{27}$, and similarly use variables $b,c,d$ (indexed appropriately) for the GIOs in types $\{udd, QdL, uuuee\}$.
A Gr\"obner basis computation shows that
\begin{equation}
d_1 = \frac{6a_{16}a_{22}-6a_{12}a_{25}}{b_9} ~,
\end{equation}
with similar expressions for $\{d_2, \cdots, d_6\}$.
This proves each of the $6$ $uuuee$ GIOs are rational functions in the GIOs of $Y_3$, and hence the map $Y_4\rightarrow Y_3$ is birational.
Choosing the ``right'' order for the projections is quite delicate, but is made possible by the extra algebraic structure of a \textit{multigrading}, described in detail in~\cite{HJNSS:2025}.

\sec{Results and Conclusion}
We prove that the master space $\cF$ and the vacuum moduli space $\cM$ of $W_{\text{minimal}}$ are complex affine algebraic varieties, each having three irreducible components.
These components intersect as in Figure~\ref{fig:two}; numbers indicate (affine) dimensions of components and intersections.
\begin{figure}[h]
\begin{center}
 \vspace{-1.3in}
\includegraphics[width=3.5in]{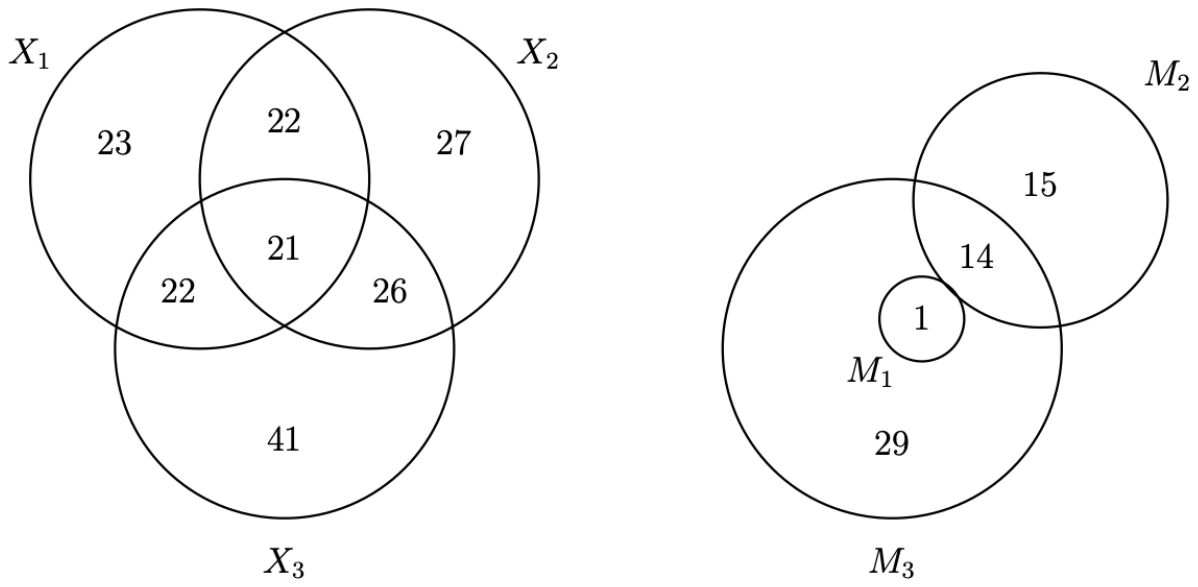} 
 \vspace{-1.7in}
\end{center}
\capt{Components and intersection dimensions.}
\label{fig:two}
\end{figure}

We determine explicit defining equations for birational models of each of these varieties and describe the corresponding birational models geometrically.
In particular, component $M_1$ is a line, and components $M_2$ and $M_3$ are rational varieties of dimension $15$ and $29$, with intersection dimensions depicted in Figure~\ref{fig:two}.
A note on terminology: a \textit{birational} map is multi-variable function which is rational (polynomial over polynomial) and whose inverse is also rational. Rationality is important because it means that on the complement of a set of measure zero (formally, on a Zariski open set), we can parameterize the points of each of the three components of the vacuum moduli space by an affine space of dimension equal to that of the component (dimension $1$, $15$, or $29$).

For example, on $M_3$ take a point on $X_3$ (which we understand explicitly in terms of equations), and use the GIOs of type $QdL$, $uude$, $udd$ to map the point to $\CC^{63}$. We have explicit polynomial equations whose vanishing defines the corresponding $29$ dimensional image of $X_3 \subseteq \CC^{63}$.
Then for any point on that image, we can lift that point to a unique point of $\CC^{973}$ which is a point of $M_3$. On the complement of a set of measure zero, every point of each of the components of the vacuum moduli space is given this way.
We also prove that restricting the component $M_2$ to the electroweak sector (hence, setting the quark variables $u=\text{up}$, $d=\text{down}$, $Q=\text{doublet}$ to zero) recovers the results of~\cite{He:2014oha} on the electroweak sector of the vacuum moduli space. 

We obtain similar results for the master space $\cF_{\text{MSSM}}$ and vacuum moduli space $\cM_{\text{MSSM}}$ of $W_{\text{MSSM}}$ when right-handed neutrinos are included. 
\begin{equation}
    \cF_{\text{MSSM}} = Z_1 \cup Z_2 \cup Z_3 ~, 
\end{equation}
where $Z_1=X_1$, $Z_3=X_3$, and $Z_2 \subseteq X_2$.
The three neutrino variables $\nu_i$ all vanish on $\cF_{\text{MSSM}}$, so we may regard $\cM_{\text{MSSM}}$ and $\cM$ as contained in the same ambient target space $\CC^{973}$.
Therefore, in the corresponding vacuum moduli spaces, the components corresponding to the images of $X_1=Z_1$ and $X_3=Z_3$ also coincide.
However, $Z_2$ is a codimension two subvariety of $X_2$ (so has measure zero in $X_2$), and as all GIO types vanish on $Z_2$, the vacuum moduli space image of $Z_2$ is zero.
The image of $Z_1$ is contained in the image of $Z_3$, so the vacuum moduli space for $W_{\text{neutrinos}}$ is irreducible and equal to $M_3$.

We summarize our findings as follows.
The vacuum moduli space of the MSSM with minimal superpotential~\eqref{eq:minimal} is the union of:
\begin{itemize}
\item $M_1$, a complex line $\mathbb{C}$, which is parametrized by a GIO of type $udd$.

\item a component $M_2$ which is birational (see definition above) to $\mathbb{C}^{15}$, parametrized by GIOS of types
$LH, LLe, QdL, dddLL, (QQQ)_4LLLe$.
The geometry is intimately related to a classical object called the Segr\`e variety.

\item a component $M_3$ which is birational (see definition above) to $\mathbb{C}^{29}$, parametrized by the vevs of GIOs of types $QdL$, $udd$, $uude$, $uuuee$, $dddLL$, $LLe$, $QQQL$, $QuQd$, $QuLe$, $QQQQu$, $QuQue$, $(QQQ)_4LLLe$, and $uudQdQd$, and contains $M_1$.
\end{itemize}

As an extra check, for the superpotential \textit{with} right-handed neutrino, decoupling the quark sector by setting $Q\!=\!u\!=\!d=\!0$, quickly produces a cone over the Veronese surface as the moduli space, in agreement with~\cite{He:2014loa}.

\sec{Closing remarks}
As noted earlier, the Gr\"obner basis algorithm is doubly exponential in the number of variables.
Algebraically, the process of computing the image of a polynomial map necessitates working in a product space, with the number of variables equal to the sum of the dimension of the source space $m=49$ and target space $n=973$.
If $d$ is the degree of the largest polynomial defining the parameterization, then 
an upper bound for the computational complexity is 
\begin{equation}
    d^{2^{n+o(n)}} ~.
\end{equation}
As there are GIOs of degree $7$, this means the potential complexity of the computation for the vacuum moduli space is $\sim 7^{2^{1023}}$, which is indeed prohibitive. 

However, symmetry is a fundamental feature of many problems arising in nature.
Our results serve as a proof of concept that, despite theoretical bounds that are astronomical, in practice methods of computational algebraic geometry can be of utility in physics.
In particular, such methods allow us to determine the vacuum moduli space.
A detailed description of our methods and results appears in~\cite{HJNSS:2025}.

Engineering a bottom up brane realization of the MSSM requires further work.
For the examples in~\cite{Douglas:1996sw}, the vacuum geometry of the D-brane gauge theory encodes the transverse directions to the worldvolume.
In order to obtain a low dimensional geometry, we can envision adding extra terms to the superpotential in a principled way and repeating the exercise outlined in this letter.
On phenomenological grounds, the additional terms will be non-renormalizable.
In the electroweak sector, including higher dimensional R-parity preserving operators preserves some of the vacuum structure~\cite{Gray:2006jb,He:2014oha,He:2015rzg}.
The string constructions of the MSSM and SUSY grand unified theories starting from the vacuum moduli space will be the subject of future work.

\sec{Acknowledgements}
YHH is supported by a Leverhulme Trust Research Project (Grant No.\ RPG-2022-145) and STFC grant ST/J00037X/3.
VJ is supported by the South African Research Chairs Initiative of the DSTI and NRF, grant 78554.
BDN is supported by NSF PHY-2209903.
HS was supported by NSF DMS-2006410 and a Leverhulme Visiting Professorship at Oxford.
MS was supported by NSF DMS-2001367 and a Simons Fellowship at Oxford.

\bibliography{refsprl}

\end{document}